\documentclass [winedt,yap]{iitparc}
\usepackage{cite}
\usepackage{amsmath,amssymb,amsfonts,bm}
\usepackage{eufrak}
\usepackage{lscape}
\usepackage{floatfig,wrapfig,epsfig}
\usepackage{subfigure}
\usepackage{color}
\usepackage{psboxit}
\usepackage{rotating}
\usepackage{curves}
\usepackage{ulem}
\usepackage[mathscr]{eucal}
\RequirePackage{psfrag}
\RequirePackage{graphicx}

\begin{document}\normalem
\initfloatingfigs
\frontmatter          
%
%
\IssuePrice{25.00}%
\TransYearOfIssue{2009}%
\TransCopyrightYear{2009}%
\OrigYearOfIssue{2009}%
\OrigCopyrightYear{2009}%
\TransVolumeNo{70}%
\TransIssueNo{3}%
\OrigIssueNo{3}%
%
\mainmatter              
%
\setcounter{page}{128}
\CRubrika{DETERMINATE SYSTEMS}
\Rubrika{DETERMINATE SYSTEMS}
%
\newtheorem*{conj*}{Conjecture}{\bfseries}{\itshape}
\title{Coordination in multiagent systems\\
       and Laplacian spectra of digraphs}

\titlerunning{Multiagent systems and Laplacian spectra}

\author{P. Yu. Chebotarev and R. P. Agaev}
\authorrunning{Chebotarev and Agaev}
\OrigCopyrightedAuthors{P.Yu. Chebotarev and R.P. Agaev}

\institute{Trapeznikov Institute of Control Sciences, Moscow, Russia}
\received{Received July 21, 2008}
\OrigPages{pp.~136--151}

\maketitle

\begin{abstract}
Constructing and studying distributed control systems requires the analysis
of the Laplacian spectra and the forest structure of directed graphs.
In this paper, we present some basic results of this analysis partially obtained 
by the present authors. We also discuss the application of these results published 
earlier to decentralized control and touch upon some problems of spectral graph theory.
\end{abstract}


\section{Introduction}

In the cooperative control of distributed multiagent systems,
the generation of control actions is decentralized.
The actions result from negotiations between agents.
As a metaphor, we can remember [1] that the musicians of
``Persimfans''\footnote[1]{Abbreviation for Perviy Simfonicheskiy
Ansambl' (First Symphony Ensemble).},
which existed in Moscow between 1922 and 1932, performed extremely complex
musical compositions \textit{without a conductor}. The string section formed
a full circle (partly with their backs to the audience), while the wind
section was situated inside of that circle. Every musician not only heard,
but was also able to see the others. This way the magic chemistry among the performers,
harmony and dynamic coordination between all the participants brought such a
synchronization, that it became the substitute of the role of the
conductor. Amazingly, one of the Persimfans' distinguishing features was their ability,
according to the testimony of the extremely demanding and tough critics, to maintain
a particularly subtle and profoundly individual approach to the interpretation of
musical pieces, normally unthinkable without the help of the conductor.

The theory of decentralized control has a long history. In addition
to the works discussed in [1], one can remember the theory of
statistical consensus by DeGroot [2], methods of step-by-step
coordination of expert judgements, such as Delphi methodology
developed at the RAND Corporation in the 1950s [3], works on
distributed networked computations and distributed decision-making
[4, 5] and, of course, modeling the collective behavior
of animals (see, e.\,g., [6]). Starting from 2003 (approximately)
we can observe an avalanche of publications on decentralized control
connected with the application of spectral graph theory in this field.

In this paper, we discuss continuous and discrete models of distributed
coordination which can be considered as basic models of decentralized control.
It is shown how the recent advantages in the algebraic graph theory (including
some results published by the present authors before the ``boom'' imploded in 2003)
can be applied in this area.

\section{A continuous model of distributed consensus}

Consider the basic continuous distributed consensus algorithm:
\begin{gather}
\dot{x}_i(t)
=-\sum\limits_{j=1}^n a_{ij}(t)\left({x_i(t)-x_j(t)}\right),\quad i=1,\dots,n.
\end{gather}

Here, $n$ is the number of agents, $x_i(t)$ is an information state
(characteristic, parameter, etc.) of the $i$th agent,
$a_{ij}(t)\ge 0$ is the weight with which agent $i$ takes into account
the discrepancy in the information state with agent $j$.
The information states can be, among others, positions (if the
agents need to rendezvous in space), velocities (if they execute decentralized
formation maneuvers), arrival times (if these must be synchronized), and so on.

Decentralized control usually requires solving more complex problems than
simply reaching a consensus. For example, if moving in
formation is considered, then a typical task is moving along a prescribed
course and in a prescribed and fixed configuration. During a violent maneuver
the configuration can be altered, but after the maneuver it must be restored.

Alteration and restoration of a prescribed geometric shape are also
typical when a ``flock'' of moving physical objects encounters an obstacle
or a hazard. It should be noted that reaching consensus is an important element
of control strategy in all such cases. Actually, it is a key element,
because it usually determines the stability of the system, its controllability, etc.
That is why the analysis and synthesis of consensus algorithms, such as (1),
is a necessary component of solving various problems of decentralized control.
In this paper we focus on graph theoretic results underlying the analysis
of distributed consensus algorithms.

Let $\Gamma(t)$ be the \textit{communication digraph\/} associated with the
consensus model~(1). The vertices of $\Gamma(t)$ are identified with the
agents, and $\Gamma(t)$ has an arc from vertex $j$ to vertex $i$
(denoted by $(j,\,i)$ or $j\to i$) if and only if $a_{ij}(t)\ne 0$.
The presence of this arc in $\Gamma(t)$ means that agent $i$
coordinates its information state with that of~$j$. The weight of the $(j,\,i)$
arc is~$a_{ij}(t)$.

In matrix form, model (1) can be written as
\begin{gather}
\dot{x}(t)=-L(t)x(t),
\end{gather}
where $x(t)=\left({x_1(t),\dots,x_n(t)}\right)^{\rm\scriptscriptstyle T}$ and the
matrix $L(t)=[\ell_{ij}(t)]_{n\times n}$ is defined as follows:
\begin{gather}
\ell_{ij}(t)=\left\{\begin{array}{ll}
{-a_{ij}(t),}                                 & {j\ne i,}\\[3mm]
\displaystyle\sum\limits_{k\ne i}{a_{ik}(t)}, &{j=i.}\
\end{array}\right.
\end{gather}

$L(t)$ is the \textit{Kirchhoff matrix\/} of the communication digraph~$\Gamma(t)$.
Sometimes, instead of $\Gamma(t)$, the digraph ${\Gamma}'(t)$ is considered
such that $j\ne i$ and $a_{ij}(t)\ne 0$ result in the presence of the
$i\to j$ arc (instead of $j\to i$ in~$\Gamma(t)$). With respect to ${\Gamma}'(t)$,
the matrix $L(t)$ defined above is the \textit{Laplacian matrix}. The classes
of Kirchhoff matrices and Laplacian matrices coincide; they only differ in the way
they are assigned to digraphs. We will call $L(t)$ the \textit{Laplacian matrix
of the algorithm\/}~(1).

The process (1) is said to be {\it convergent\/} if for any initial conditions
$x_i(0),\:i=1,\dots,n,$ and every $i,j=1,\dots,n$, it holds that
$\,\left|{x_i(t)-x_j(t)}\right|\to 0\,$ as $\,t\to\infty$.

The convergence properties of the algorithm (1) are determined by the spectral
properties of the matrix $L(t)$.

\section{Simple properties of the Laplacian matrices}

By definition, $L(t)$ has zero row sums, consequently, it is singular and the
vector ${\mathbf{1}}=[1,\dots,1]^{\rm\scriptscriptstyle T}$ belongs to its kernel.
Since the off-diagonal entries of $L(t)$ are nonpositive, the diagonal
entries being nonnegative, and $L(t)$ has a weakly dominant diagonal,
by Ger\v{s}gorin's, theorem the real parts of all nonzero eigenvalues of $L(t)$
are strictly positive. That is why all nonzero eigenvalues of the matrix
$(-1)\!\cdot\! L(t)$ in (2) have strictly negative real parts.

The Laplacian matrix of an undirected graph is symmetric,
positive semidefinite, and its spectrum is real and nonnegative.
Moreover, $0$ is a simple eigenvalue of this matrix if and only if the
corresponding graph is connected (see, e.\,g., [7]). If
$0=\lambda_1\le\lambda_2\le\ldots\le\lambda_n$ are the Laplacian eigenvalues of
an undirected graph, then $\lambda_2$ is referred to as the \textit{algebraic
connectivity\/} of the graph. This concept was introduced by M.\,Fiedler, and it is
widely used in both theoretical studies and applications.

The Laplacian matrices of undirected graphs have been much studied. They date back
to the famous matrix tree theorem by Kirchhoff; A.K.\,Kelmans pioneered their
systematic investigation in the 1960s and 1970s; in the 1990s, the results of their study
were presented in a series of reviews by R.\,Merris ([7] is one of them) and B.\,Mohar;
then the monographs [8,\,9] were published.

The Laplacian matrices of directed graphs and relationships between their
properties and the properties of the corresponding digraphs are still very poorly studied.
To some extent, this is due to the fact that the mathematical problems which involve
the complex spectra of digraph Laplacians are much more difficult than the corresponding
problems regarding the real spectra of ordinary graphs. At the same time, the need for
such studies is well recognized. Distributed control is one of the applications where
this need is particularly urgent.

\section{The convergence of consensus algorithms}

Assume that the Laplacian matrix $L(t)$ is kept constant: $L(t)=L$.
It is not difficult to formulate a matrix-theoretic necessary and sufficient condition of
convergence for the algorithm~(2). As already mentioned, the vector
${\mathbf{1}}=[1,\dots,1]^{\rm\scriptscriptstyle T}$ belongs to the kernel of~$L$. If $0$ is a simple
eigenvalue of $L$, then $x(t)\to\bar{x}\,{\mathbf{1}}$, where $\bar{x}$ is a scalar,
consequently, $\left|x_i(t)-x_j(t)\right|\to 0$ as $t\to\infty$ for all $i,j=1,\dots,n$, and
the coordination trajectories converge. As shown in [10, Theorem~4], the zero eigenvalue of $L$
is always \textit{semisimple\/}, i.\,e., its algebraic and geometric multiplicities coincide.
Hence, if this eigenvalue is not simple, then the kernel of $L$ is not one-dimensional, and
the convergence is violated.

Thus, the convergence analysis of algorithm (2) reduces to the determination of
conditions under which $0$ is a simple eigenvalue of~$L$.

\section{Ranks of Laplacian matrices and a convergence criterion for distributed
         consensus algorithms}

The rank of the digraph Laplacian was studied in~[11].

Recall some graph theory notation. A digraph is called \textit{strongly connected\/}
(or {\it strong}) if it contains directed paths from every vertex into every other vertex.
Every maximal (by inclusion) strong subgraph of a given digraph is called its
\textit{strong component\/} or \textit{bicomponent}. A~\textit{basis bicomponent\/} of a
digraph is a bicomponent such that the digraph does not have any arcs flowing into this
bicomponent from outside. It is easy to verify that every vertex of a digraph is reachable
by a path from at least one basis bicomponent. If a basis bicomponent consists of a single
vertex, we call it an \textit{undominated\/} vertex of the digraph. A digraph is \textit{weakly
connected\/} if the graph obtained from it by replacing all arcs with undirected
edges is connected. \textit{Weak components\/} of a digraph are the maximal by inclusion
weakly connected subgraphs of this digraph. A digraph is \textit{unilaterally connected}
if for any vertices $i$ and $j\ne i$, it contains either a directed path from $i$ into
$j$ or a directed path from $j$ into $i$ (or both). \textit{Unilateral components\/} of a
digraph are its maximal by inclusion unilaterally connected subgraphs. It should be noted that
the relation of unilateral connectivity need not be transitive, so it does not induce a
decomposition of the vertex set into equivalence classes. Therefore, as distinct from the
strong and weak components, unilateral components can overlap.

A subgraph of a digraph is {\it spanning\/} if the vertex sets of the graph and
subgraph coincide. A~\textit{diverging tree\/} is a rooted directed tree containing
directed paths from the root into all other vertices. A~\textit{diverging forest\/} is a
rooted directed forest all of whose weak components are diverging trees.

For an arbitrary digraph $\Gamma$, consider its spanning diverging forests.
Such forests are also called \textit{out-forests\/} of~$\Gamma$.
A spanning diverging forest $F$ of $\Gamma$ is a \textit{maximum out-forest\/}
if $\Gamma$ has no out-forest with the number of arcs greater than in~$F$.
It is easily seen that every maximum out-forest has the minimum possible number of weak
components (diverging trees); this number will be called the \textit{out-forest dimension\/}
of~$\Gamma$ and denoted by~$d$. The number of arcs in any maximum out-forest is obviously
$n-d$, where $n$ is the number of vertices in~$\Gamma$. The following results were obtained
in~[11].

\begin{proposition}
Let $L$ be the Kirchhoff matrix of a digraph~$\Gamma$. Then $\rank\,L=n-d,$ where $n$ is the
number of vertices in\/ $\Gamma$ and $d$ is the out-forest dimension of\/~$\Gamma$.
\end{proposition}

\begin{proposition}
The out-forest dimension of a digraph is equal to its number of basis bicomponents.
\end{proposition}

\begin{proposition}
The out-forest dimension of a strong digraph is unity.
\end{proposition}

\begin{proposition}
The out-forest dimension of a digraph is no less than its number of weak components and does
not exceed the number of its strong components and the number of its unilateral components.
\end{proposition}

Zero is a simple eigenvalue of $L$ if and only if $\rank\,L=n-1$. That is why the above
propositions imply Corollary~1.

\begin{corollary}
Let $L$ be the Kirchhoff matrix of a digraph~$\Gamma$. Then $0$ is a simple eigenvalue
of $L$ if and only if\/
$\Gamma$ has a spanning diverging tree or$,$ equivalently$,$
$\Gamma$ has only one basis bicomponent.
\end{corollary}

By Propositions~1 and 3, the premise of Corollary~1 is satisfied, for instance, for
any strong digraph. Later on, the first statement of Corollary~1 was obtained in [12--16].

Corollary 1 provides a convergence criterion for the consensus algorithm~(2).

\begin{theorem}
The consensus algorithm~$(2)$ with a stable Laplacian matrix $L(t)=L$
converges to a vector with equal components
for any vector of initial conditions $x(0)$
if and only if the corresponding communication digraph\/
$\Gamma$ has a spanning diverging tree or$,$ equivalently$,$
has a unique basis bicomponent.
\end{theorem}

A more general problem is describing the whole domain in the space of initial conditions
belonging to which guarantees the fulfillment of the
$\left|{x_i(t)-x_j(t)}\right|\to 0$, $i,j=1,\dots,n,$ condition, provided that
convergence does not generally hold. This problem reduces to the analysis of the kernel of~$L$.

\section{The kernel and eigenprojection of the digraph Laplacian}

The \textit{eigenprojection\/}\footnote[2]{It is also called \textit{principal idempotent}.}
corresponding to eigenvalue $0$ or, for short, simply \textit{eigenprojection\/} of a square
matrix $A$ is a projection (i.\,e., an idempotent matrix) $Z$ such that
$R(Z)=N(A^{\nu})$ and
$N(Z)=R(A^{\nu})$, where $R(A)$ and $N(A)$ are the range and the kernel
(null space) of $A$, respectively, and $\nu=\mathrm{ind}\,A$ is the index of $A$, i.\,e.,
the minimum $k\in\{0,\,1,\dots\}$ such that $\rank\,A^{k+1}=\rank\,A^k$. In other words, the
eigenprojection of $A$ is the projection on $N(A^{\nu})$ along $R(A^{\nu})$.
The eigenprojection is unique, as an idempotent matrix is uniquely determined by its range
and kernel. A~number of equivalent definitions of eigenprojection can be found in~[17].

In [11] we considered the \textit{normalized matrix of maximum out-forests of a digraph\/}
denoted by $\bar{J}$ (the definition of this matrix is given below in Section~8) and
proved the following properties of~$\bar{J}$.

\begin{proposition}
Let $L,$ $d$ and $\bar{J}$ be the Kirchhoff matrix$,$ the out-forest dimension and the
normalized matrix of maximum out-forests of digraph~$\Gamma,$ respectively. Then
$\rank\bar{J}=d;$ $\;\bar{J}^2=\bar{J};$ $\;L\bar{J}=\bar{J}L=0$.
\end{proposition}

By the second statement of Proposition~5, $\bar{J}$ is a projection.
By the third statement, $R(\bar{J})\subseteq N(L)$ and $R(L)\subseteq N(\bar{J})$.
Using the relationship between the ranks of $\bar{J}$ and $L$
($\rank\bar{J}=d$ and $\rank\,L=n-d$ by Proposition~1) and the semisimplicity of $0$ as
the eigenvalue of $L$, we obtain that the above inclusions can be replaced with equalities.
Finally, the semisimplicity of $0$ as the eigenvalue of $L$ implies that
${\mathrm{ind}}\,L=1$. Thus, the following two propositions hold true.

\begin{proposition}
Let $L$ and $\bar{J}$ be the Kirchhoff matrix and the normalized matrix of maximum
out-forests of $\Gamma,$ respectively. Then $\bar{J}$ is the eigenprojection of~$L$.
\end{proposition}

\begin{proposition}
The linear spans of the columns and the rows of $\bar{J}$ coincide with the kernel
and the left null space of $L,$ respectively.
\end{proposition}

By virtue of Propositions~6 and 7, matrix $\bar{J}$ is useful for the analysis of
continuous distributed consensus algorithms. In Section~7, we will see that $\bar{J}$
is also applicable to the study of iterative consensus algorithms.

\section{Iterative consensus algorithms}

Consider the finite-difference counterpart of the continuous consensus algorithm (1) with
constant coefficients $a_{ij}$:
\begin{gather}
x_i(k+1)
=x_i(k)-\varepsilon\sum\limits_{j=1}^n{a_{ij}\left({x_i(k)-x_j(k)}\right)},
\quad i=1,\dots,n,
\end{gather}
where $k$ is the discrete time and $\varepsilon>0$ is the step size.
Rewrite algorithm (4) in matrix form:
\begin{gather}
x(k+1)=P\,x(k),
\end{gather}
where
\begin{gather}
P=I-\varepsilon L.
\end{gather}

If the step size $\varepsilon$ is small enough, then $P$ is a row stochastic matrix.
This follows from the fact that $L$ has nonpositive off-diagonal entries and zero row sums.
The corresponding condition of the ``smallness'' of $\varepsilon$ is as follows~[11]:
\begin{gather}
0<\varepsilon\le\left(\max_i\sum\limits_{j\ne i}
a_{ij} \right)^{-1}.
\end{gather}

The matrix $P=\exp(-\varepsilon L)$ corresponds to the continuous consensus algorithm (2),
and (6) can be considered as the expansion of $\exp(-\varepsilon L)$ to the linear term.
Matrix (6) is sometimes called the \textit{Perron matrix with parameter\/} $\varepsilon$ of
digraph~$\Gamma$.

From (5), for any natural $m$ one has
\begin{gather}
x(m)=P^mx(0),
\end{gather}
therefore, the properties of the process (5) are determined by the properties of the
sequence $\{P,P^2,\dots,P^m,\dots\}$. From the theory of Markov chains it is known
that this sequence need not converge (the necessary and sufficient condition of its
convergence is the aperiodicity of the chain), but the Ces\`aro limit (also called the time
average limit)
\begin{gather}
P^{\infty} =\lim_{m\to \infty } \frac{1}{m}\sum\limits_{i=1}^m{P^i},
\end{gather}
always exists and coincides with the limit of the sequence $\{P,P^2,\dots,P^m,\dots\}$ in
case the latter converges. Otherwise, if the chain is periodic with period $s$, then
\begin{gather*}
P^{\infty}=\frac{1}{s}\left(P^{(1)}+\ldots+P^{(s)}\right)
\end{gather*}
where $P^{(1)},\dots,P^{(s)}$ are the limits of its converging subsequences:
$P^{(i)}=\lim\limits_{j\to \infty}P^{js+i}$.

It was shown in [18] that $P^{\infty}$ is the eigenprojection of $I-P$.
Hence, by (6), $P^{\infty}$ is the eigenprojection of~$L$ as well. This fact along
with the uniqueness of the eigenprojection and Proposition~6 imply Proposition~8.

\begin{proposition}
Let a row stochastic matrix $P$ be connected with the Kirchhoff matrix $L$ of a
digraph $\Gamma$ by equation $(6)$ with $\varepsilon>0$.
Then the matrix $P^{\infty}$ defined by $(9)$ coincides with the normalized matrix
$\bar{J}$ of maximum out-forests of\/~$\Gamma$.
\end{proposition}

Thus, the matrix $P^{\infty}$, which determines the asymptotic behavior of the consensus
algorithm (4), is equal to the normalized matrix of maximum out-forests of the communication
digraph that corresponds to this algorithm. The matrix $\bar{J}$ is thereby important
for the analysis of iterative consensus processes. Indeed, according to (8),
to know the average asymptotic state of the algorithm (4), it suffices to consider the
product $P^{\infty}x(0)=\bar{J}x(0).$
To compute the matrix $\bar{J}$, one can use the algorithm proposed in [19,~20],
which reduces to $\min\{n-d-1,\;0\}$ multiplications of matrices of order~$n$.

It should be remarked that Proposition 8 coincides with the Markov chain tree theorem first
obtained by Wentzell and Freidlin [21] and rediscovered by Leighton and Rivest~[22].

Consider the convergence of the consensus algorithm~(4). Since the spectral radius of $P$ is
$1$, the convergence is violated only if $P$ has an eigenvector not proportional to
${\mathbf{1}}=[1,\dots,1]^{\rm\scriptscriptstyle T}$ and corresponding to an eigenvalue of modulus~1.
The subspace of invariant vectors of $P$ coincides with the kernel of $L$, hence it is
one-dimensional if and only if the premise of Theorem~1 holds. As for the complex
eigenvalues of modulus~1, $P$ can have them \textit{only if\/} $\varepsilon$ coincides with
the right endpoint of the interval~(7). Indeed, in the opposite case, the increase
of $\varepsilon$ up to the endpoint of the interval would have resulted in the appearance
of an eigenvalue with the modulus greater than~1, in contradiction with the stochasticity
of $P$ (see also~[23]).

Consequently, the convergence of consensus algorithm~(4) is guaranteed by the fulfillment
of the premise of Theorem~1 together with the strict form of inequality~(7). Otherwise, if
$\varepsilon$ coincides with the right endpoint of the interval~(7) and the Markov chain
determined by the matrix $P=I-\varepsilon L$ is periodic, then the convergence is violated.

\section{The normalized matrix of maximum out-forests}

Can the entries of the matrix $\bar{J}$, which proves to be useful for the analysis of consensus
algorithms, be interpreted in terms of the communication digraph $\Gamma$\,?
Such a connection is specified by the very definition of $\bar{J}$:
every entry $\bar{J}_{ij}$ of $\bar{J}$ is defined [11] as the ratio of the total
weight\footnote[3]{The weight of a digraph is the product of the weights of all its arcs.}
of $\Gamma$'s maximum out-forests that have vertex $i$ belonging to a tree diverging from $j$
to the total weight of all maximum out-forests in~$\Gamma$.

The following theorems [11] summarize the properties of~$\bar{J}$.

\begin{theorem}
Suppose that $V(\Gamma)=\{1,\dots,n\}$ and $E(\Gamma)$ are the vertex set and arc set of a
digraph\/ $\Gamma,$ $K$ is the vertex set of some basis bicomponent of\/ $\Gamma,$ $K^+$ is
the set of vertices reachable by paths from $K$ and unreachable from the other basis
bicomponents of\/ $\Gamma,$ $\widetilde{K}$ is the union of the vertex sets of all basis bicomponents$,$
$d$ is the out-forest dimension of\/ $\Gamma,$ $\bar{J}$ is the normalized matrix of
maximum out-forests of\/ $\Gamma,$ ${\mathsf{T}}$ is the set of spanning diverging trees in
the restriction\/ $\Gamma_K$ of\/ $\Gamma$ to $K,$ ${\mathsf{T}}^j$ is the subset of\/
${\mathsf{T}}$ consisting of the trees that diverge from~$j,$ $\Gamma_{-K}$ is the
spanning subgraph of\/ $\Gamma$ with the edge set $E(\Gamma)\setminus E(\Gamma_K),$
${\mathsf{P}}^{K\to i}$ is the set of maximum out-forests of\/ $\Gamma_{-K}$ in which
$i$ is reachable from some vertex that belongs to~$K,$ $\varepsilon(\cdot)$ designates the
weight of a set of subgraphs $($equal to the sum of the weights of its elements$),$
$\sigma_{n-d}$ is the weight of the set of maximum out-forests of\/~$\Gamma$.
Then the following statements are true$:$

$1$.~$\bar{J}$ is a row stochastic matrix\textup:
$\bar{J}_{ij}\ge 0,\;$ and $\;\sum\limits_{k=1}^n
{\bar{J}_{ik}=1}$ $\:i,$ $j$ $\in V(\Gamma)$\textup;

$2$.~$\bar{J}_{ij}\ne 0\;\Leftrightarrow$ $(j\in \widetilde{K}$ and $i$ is reachable
from $j$ in~$\Gamma)$\textup;

$3$.~Let $j\in K$. For any $i\in V(\Gamma),$
$\bar{J}_{ij}={\varepsilon({\mathsf{T}}^j)\,
               \varepsilon({\mathsf{P}}^{K\to i})}/{\sigma_{n-d}}$.
Moreover$,$ if $i\in K^+,$ then
$\bar{J}_{ij}=\bar{J}_{jj}={\varepsilon({\mathsf{T}}^j)}/
                                     {\varepsilon({\mathsf{T}})}$\textup;

$4$.~$\sum\limits_{j\in K}{\bar{J}_{jj}}=1$. In particular$,$
 if $j$ is an undominated vertex$,$ then $\bar{J}_{jj} =1;$

$5$.~If $j_1,j_2\in K,$ then $\bar{J}_{\cdot j_2 }=
\left({{\varepsilon({\mathsf{T}}^{j_2 })}/
       {\varepsilon({\mathsf{T}}^{j_1 })}}\right)\bar{J}_{\cdot j_1 },$ where
$\bar{J}_{\cdot j_1}$ and $\bar{J}_{\cdot j_2}$ are the $j_1$ and $j_2$ columns
of~$\bar{J}$.
\end{theorem}

\begin{corollary}[from the statement 3 of Theorem 2]$\ $

$1$. The normalized matrix $\bar{J}_K
=\left[\bar{J}_{ij}^K \right]$ of maximum out-forests of the digraph~$\Gamma_K$
coincides with the principal submatrix of $\bar{J}$ corresponding to the basis
bicomponent~$K$.

$2$.~If $i\in K^+$ and $j\in K^+\backslash K,$ then $\bar{J}$ is preserved under the
variations of the weight of $(i,\,j)$.
\end{corollary}

The following Theorem~3 is concerned with the comparison of the entries of~$\bar{J}$.

\begin{theorem}
In the notation of Theorem~$2,$ let $K(k)$ be a basis bicomponent of\/ $\Gamma$
that contains a vertex $k\in\widetilde{K}$. For all $i,j\in V(\Gamma)$ we have\textup:

$1$.~$\bar{J}_{ii}\ge\bar{J}_{ji}$\textup;

$2$.~If $\bar{J}_{ii}>\bar{J}_{ji},$ then $i\in\widetilde{K},$
$j\notin K^+(i),$ therefore$,$ $\Gamma$ contain no paths from $j$ into~$i$\textup;

$3$.~If $\bar{J}_{ii}>\bar{J}_{ji}>0,$ then $j\notin\widetilde{K},$
consequently$,$ $j$ is not the root in any maximum out-forest of\/~$\Gamma$\textup;

$4$.~If $\bar{J}_{ij}>0,$ then $\bar{J}_{ii}=\bar{J}_{ji}$.
\end{theorem}

As mentioned above, these properties are useful for the analysis of the consensus algorithms
(2) and (5) as well as generalizations of these algorithms, in particular, when the convergence
is not guaranteed by the structure of the communication digraph. Some of the foregoing
results regarding the kernel of the digraph Laplacian were also obtained in~[24].

\section{Other problems of decentralized control}

The distributed consensus algorithm (2) is the simplest algorithm of decentralized control.
Yet this algorithm is a basic one, because the linear operator of agreement
$(-1)\!\cdot\! L(t)$ it contains is an indispensable constituent of more complex algorithms as well.
Therefore, the convergence properties of these more complex algorithms are in many respects
determined by the spectral properties of the corresponding Laplacian matrices and by the forest structure
of the related communication digraphs.

In this section, we briefly survey some modifications of the consensus algorithm (2) and some
other related problems of decentralized control.

First of all, it should be noted that the model (2) does not generally imply that the
communication digraph $\Gamma(t)$ is fixed. The rejection of this assumption makes the
model more realistic. Indeed, in many applications the agents primarily communicate with
the closest neighbors, but those can be acquired or lost during the motion. To preserve the
relative simplicity of the model, some restrictions on the structure of the communication
digraph are applied. Usually, either it is supposed to be piecewise constant and a set of
possible switchings is specified or the weights $a_{ij}(t)$ are defined with a permissible
variation. Sometimes fluctuation of the weights is described by a probabilistic model.
As a result, the analysis, in most cases, reduces to solving a series of problems with
stable communication digraphs and combining the partial solutions. In some cases, the analysis
involves the study of infinite products of stochastic matrices taken from a specific
set~[25].

Besides that, the model becomes more realistic if it allows communication delays.
These delays can be associated with either information transmission or with processing messages
after receipt. In this case, the model of the form (1) contains terms such as
$x_j(t-\delta_{ij})$, where $\delta_{ij}$ is a delay typical of the pair of agents
$(i,\,j)$.
If the delays are the same for all pairs of agents, then their presence does not essentially
alter the properties of the coordination trajectories~[26]. More general cases were considered in~[27].

In many technical applications it is desirable to maximize the rate of convergence. An
approach to solving such problems was developed in~[28]. It is noteworthy that in case
of undirected graphs, the problem of the best convergence is connected [29] with the
maximization of the algebraic connectivity (the second smallest eigenvalue of the Laplacian
matrix already mentioned in Section~3) of the communication graph. The algebraic connectivity
itself can serve as a good measure of the convergence rate of the algorithm~[30]. For the
case of directed graphs, different generalizations of the algebraic connectivity were proposed,
including the second smallest real part of the Laplacian eigenvalues~[31], the second smallest
modulus of the Laplacian eigenvalues, and the second smallest eigenvalue of the symmetric
part $(L+L^{\rm\scriptscriptstyle T})/2$ of the Laplacian matrix [23,\,32]. However, in the
general case of digraphs, the problem of optimizing the convergence rate is still insufficiently
studied. Not even the problem of localizing the spectra of nonsymmetric Laplacian matrices (see
Section~10) has been completely investigated.

Let us turn to some extensions of the model (2) that have a more complex structure.
For the analysis of synchronization in systems of nonlinear oscillators and for solving
some other problems, the following class of models was considered:
\begin{gather}
\dot{x}_i=
f(x_i)-\gamma\sum\limits_{j=1}^n{a_{ij}(t)\,(x_i-x_j)},\quad i=1,\dots,n,\quad\gamma>0.
\end{gather}
It has been shown in [33] that in case of undirected communication graphs, the corresponding
synchronization process is essentially determined by the algebraic connectivity of the graph.

The consensus algorithm for a double-integrator dynamics has the form:
\begin{gather}
\ddot{x}_i=
-\sum\limits_{j=1}^n{a_{ij}(t)\left({(      x_i-      x_j)+
                               \gamma(\dot{x}_i-\dot{x}_j)}\right)\,} ,\quad i=1,\dots,n,
\end{gather}
where $\gamma>0$ controls velocity matching. It was shown in [34] that $\Gamma$
is of considerable importance here, however, for the algorithm to converge, as well as for the
basic model (Theorem~1), the communication digraph must have a spanning diverging tree.

More general differential models of the second order were used to describe the motion of the
complexes (flocks, schools, troop, swarms, etc.) of physical objects. ``Flocking'' requires both
maintaining the pattern formed by the objects' positions (up to translation and/or rotation)
and formation maneuvering [31,\,35]. In this case, the existence of a spanning diverging
tree in the communication digraph is also a necessary condition of the effective control for
many typical problems.

\section{Localizing the spectra of nonsymmetric Laplacian matrices}

For the analysis of models generalizing simple consensus models (1) or (4) it is necessary
to know the whole Laplacian spectrum of the communication digraph rather then the
sole multiplicity of its zero eigenvalue. Another important but rather difficult problem is
recovering connections between the Laplacian spectrum of a digraph and the topological
properties of the same digraph, including the ``degree of cyclicity'' and many others.

The problem of localizing the spectra of Laplacian matrices was partially solved in~[10].
Let us consider some results of this paper.

A real square matrix of order $n$ will be called a \textit{standardized\/}
\textit{Laplacian matrix\/} if
(1)~its row sums are equal to~0 and
(2)~its off-diagonal elements are nonpositive and do not exceed $1/n$ in absolute value.
This standardization enables one to consider results on the spectra of Laplacian matrices
with various~$n$ in the same framework. Standardized Laplacian matrices will be denoted by
${\widetilde L}$. If the class $G_b$ of weighted digraphs with positive arc weights not exceeding
$b>0$ is considered and $L(\Gamma)$ is the Laplacian matrix of a weighted digraph $\Gamma\in G_b$
on $n$ vertices, then the {\it standardized Laplacian matrix associated with\/ $\Gamma$ in this
class\/} is, by definition, ${\widetilde L}(\Gamma)=(nb)^{-1}L(\Gamma).$

Let $J\in{\mathbb R}^{n\times n}$ be the matrix with all entries $1/n$; $K=I-J$.
Then $K$ is the standardized Laplacian matrix of the complete digraph with all arc weights $b$ in the class~$G_b $.
Define the matrices
\begin{gather}
P={\widetilde L}+J
\end{gather}
and
\begin{gather}
\widetilde L_c=K-{\widetilde L}.
\end{gather}

By (12) and (13), $P$ is a stochastic matrix, $\widetilde L_c$ being the
standardized Laplacian matrix of the complementary weighted
digraph $\Gamma_c$ in which $(b-\varepsilon_{ij})$ is the weight of
arc $(i,j)$, $j\ne i$, provided that $\varepsilon_{ij}$ is the weight of this
arc in~$\Gamma$. If $\varepsilon_{ij}=b$, then $\Gamma_c$ has no
$(i,j)$ arc and vice versa: if $\Gamma$ has no $(i,j)$ arc,
then the weight of $(i,j)$ in $\Gamma_c$ is~$b$. It follows from
(12), (13) and the definition of $K$ that
\begin{gather}
P=I-\widetilde L_c.
\end{gather}

Consider the results [10] connecting the spectra of $\widetilde L$, $P$ and $\widetilde L_c$.

\begin{theorem}
Let ${\widetilde L}$ be a standardized Laplacian matrix$;$ let $P$ and $\widetilde L_c$
be defined by $(12)$ and $(13),$ respectively. Then for $\lambda\notin\{0,\,1\},$
the following statements are equivalent\textup:\footnote[4]{$\mathrm{sp}A$ is the spectrum of $A$.}
\begin{gather*}
\textup{(a)}~\lambda  \in{\mathrm{sp}}\,{\widetilde L};\quad
\textup{(b)}~\lambda  \in{\mathrm{sp}}\,P;\quad
\textup{(c)}~1-\lambda\in{\mathrm{sp}}\,\widetilde L_c,
\end{gather*}
and these eigenvalues have the same geometric multiplicity. Furthermore$,$ $v$ is an eigenvector of
${\widetilde L}$ corresponding to $\lambda\notin\{0,\,1\}$ if and only if the vector\,\footnote[5]{For
simplicity, we sometimes write $A/\alpha$ instead of $(1/\alpha)A$, where $A$ is a matrix and $\alpha\ne 0$
is a complex number.}
\begin{gather}
x=\left( {I-\frac{J}{1-\lambda}}\right)\,v
\end{gather}
is an eigenvector of $P$              corresponding to $\lambda$ and
   an eigenvector of $\widetilde L_c$ corresponding to $1-\lambda$.
\end{theorem}

\begin{theorem}
Let $f_{\widetilde L}  (\lambda),$
    $f_{P}             (\lambda) $ and
    $f_{\widetilde L_c}(\lambda) $
be the characteristic polynomials of $\widetilde L,$ $P$ and $\widetilde L_c,$ respectively.
Then for all $\lambda\not\in\{0,1\},$
\begin{eqnarray}
  f_{P}             (\lambda) &=& {\lambda-1\over\lambda}f_{\widetilde L}(\lambda),\\
  f_{\widetilde L_c}(\lambda) &=& (-1)^{n-1}{\lambda\over1-\lambda}f_{\widetilde L}(1-\lambda).
\end{eqnarray}
\end{theorem}

\begin{theorem}
For any standardized Laplacian matrix
${\widetilde L}$ and the corresponding stochastic matrix~$P$ defined by $(12),$
${\widetilde L}$ and $P$ are semiconvergent.
\end{theorem}

For completeness, the following theorem contains some results mentioned above.

\begin{theorem}
Suppose that $d$ and $d_c$ are the in-forest dimensions of the digraph whose standardized Laplacian matrix is
${\widetilde L}$ and the complementary digraph$,$ respectively$;$ $m_A (\lambda)$ is the algebraic
multiplicity of $\lambda\in{\mathrm{sp}}\,A;$ $V_A (\lambda)$ is the eigenspace of $A$ corresponding to~$\lambda$.
Then

\begin{tabular}{rll}
  {\rm(i)} &$m_{\widetilde L}(0)=d,$     &$m_{\widetilde L}(1)=d_c-1;$\\
 {\rm(ii)} &$m_P (0)=d-1,$               &$m_P (1)=d_c;$\\
{\rm(iii)} &$m_{\widetilde L_c}(1)=d-1,$ &$m_{\widetilde L_c}(0)=d_c,$\\
           &\multicolumn{2}{l}{and these eigenvalues are semisimple\textup;}\\
 {\rm(iv)} &\multicolumn{2}{l}
            {if $v\in        V_{\widetilde L}  (0)$ and $\:K v\ne0,$ then $\:K v\in V_P(0)=V_{\widetilde L_c}(1);$}\\
           &\multicolumn{2}{l}
            {if $x\in V_P(1)=V_{\widetilde L_c}(0)$ and $\:K x\ne0,$ then $\:K x\in        V_{\widetilde L  }(1) $.}
\end{tabular}
\end{theorem}

Let $\widetilde\Lambda_n$ be the class of standardized Laplacian matrices of order~$n$.
We now turn to the problem of localizing the spectra of the matrices $\widetilde L\in\widetilde\Lambda_n$.
The following results were obtained in~[10].

\begin{theorem}
All eigenvalues of standardized Laplacian matrices of order~$n$ belong to the meet of\/$:$

$\cbd$ two closed disks$,$
      one centered at $1/n,$
the other centered at $1-1/n,$
each having radius    $1-1/n,$

$\cbd$ two closed smaller angles$,$
one       bounded with the two half-lines drawn from $1$ through
                               $e^{-2\pi\mathrm{i}/n}$              and $e^{2\pi\mathrm{i}/n},$
the other bounded with the     half-lines drawn from $0$ through
                               $e^{-({\pi\/2}-{\pi\/n})\mathrm{i}}$ and $e^{({\pi\/2}-{\pi\/n})\mathrm{i}},$ and

$\cbd$ the band\: $|\Im(z)|\le\frac{1}{2n}\cot\frac{\pi}{2n}.$
\end{theorem}

\begin{remark*}
Using the theorem by Dmitriev and Dynkin [36] on the spectra
of stochastic matrices, it can be shown that ${\mathrm{sp}}\,{\widetilde L}$
contains an eigenvalue with argument $\frac{\pi}{2}-\frac{\pi}{n}$ if and only if
$\Gamma$ is a Hamiltonian cycle on $n$ vertices. In this case, such an eigenvalue $\lambda$ is unique,
$|\lambda|\le\frac{2}{n}\sin\frac{\pi}{n}$, and $\Im(\lambda)\le\frac{1}{n}\sin\frac{2\pi}{n}$.
The components of any corresponding eigenvector are the vertices of a regular polygon.
Similarly, ${\mathrm{sp}}\,{\widetilde L}$ contains an eigenvalue
that belongs to the segment $[1,e^{{2\pi\mathrm{i}/n}}]$ if and only if the complementary digraph $\Gamma_c$
is a Hamiltonian cycle on $n$ vertices. As above, such an eigenvalue $\lambda'$ is unique and
$\Im(\lambda')\le\frac{1}{n}\sin\frac{2\pi}{n}$.
\end{remark*}

This remark and Theorem~8 are illustrated by Fig.~1.

\begin{figure}[t]
\centering{\includegraphics[clip]{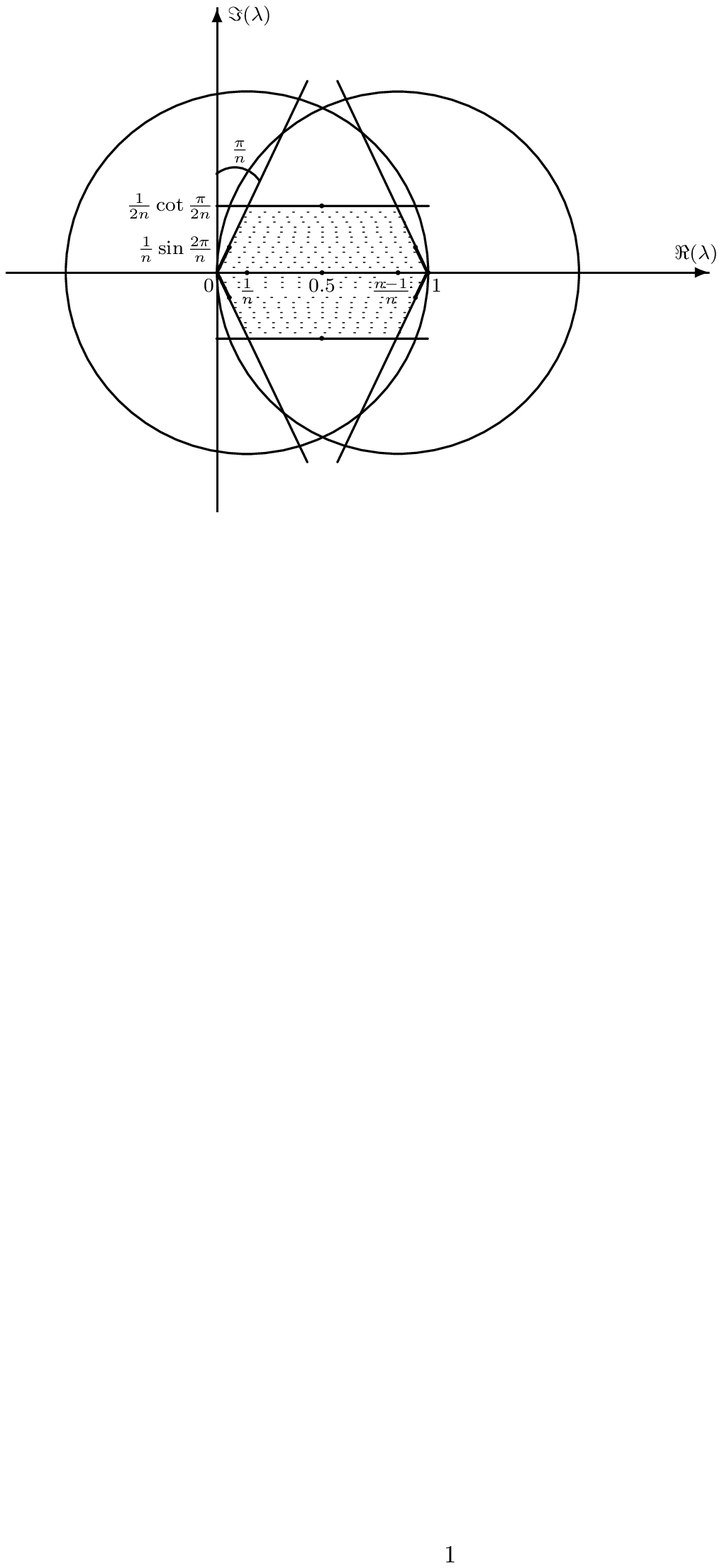}}
\caption{The domain which, according to Theorem~8, contains the spectrum of each standardized Laplacian
matrix of order~$n$ (hatched); in this figure, $n=7$.}
\bigskip
\end{figure}

Let
\begin{eqnarray}
\lambda_k(n)&=&n^{-1}\left(k-e^{-2\pi\mathrm{i}/n}-\ldots-e^{-2k\pi\mathrm{i}/n}\right)\\
\nonumber
            &=&n^{-1}\left(k-\frac{\sin\tfrac{k\pi}{n}}{\sin\tfrac{\pi}{n}}e^{-(k+1)\pi\mathrm{i}/n}\right), \quad
k=1,\dots,n-1;
\end{eqnarray}
by $S(n)$ we denote the closed convex polygon with vertices
\begin{gather}
\lambda_{0}(n)=0,                      \quad
\lambda_{1}(n),\dots,\lambda_{n-2}(n), \quad
\lambda_{n-1}(n)=1,                    \quad
\overline{\lambda}_{n-2}(n),\dots,\overline{\lambda}_1 (n).
\end{gather}

\begin{theorem}
Every point of the polygon $S(n)$ is an eigenvalue of some standardized Laplacian matrix
${\widetilde L}\in\widetilde\Lambda_n$.
\end{theorem}

Let $h(n)=\sup\{\Im(\lambda):\lambda$ is an eigenvalue of some ${\widetilde L}\in\widetilde\Lambda_n\}$.

By Theorem 8, for all $n=2,3,\dots\;$ it holds that $h(n)\le\dfrac{1}{2n}\ctg\dfrac{\pi}{2n}$.

\begin{theorem}
If $n$ is odd$,$ then
\begin{gather*}
h(n)=\frac{1}{2n}\ctg\frac{\pi}{2n},
\end{gather*}
moreover$,$ $h(n)=\Im\left(\lambda_{(n-1)/2}(n)\right),$ where
                        $\lambda_{(n-1)/2}(n)$ is defined by $(18)$.
\end{theorem}

\begin{proposition}
If $n>2$ is even$,$ then
\begin{gather*}
\max_{0\le k\le 2n-1}\Im(\lambda_k(n))
=\Im(\lambda_{n/2}(n))
=\frac{1}{n}\cot\frac{\pi}{n}<\frac{1}{2n}\cot\frac{\pi}{2n}.
\end{gather*}
\end{proposition}

\begin{corollary*}[from Theorems~8 and 9 and Proposition~9]
$\lim\limits_{n\to\infty}h(n)=1/\pi$.
\end{corollary*}

The following conjecture is yet unproved.

\begin{conj*}
All eigenvalues of standardized Laplacian matrices of order $n$
belong to the polygon~$S(n)$ whose vertices are defined by~$(19)$.
\end{conj*}

Note that the vertices $\lambda_k(n)$ and $\overline{\lambda}_k(n)$ of the polygon $S(n)$ belong to the spectrum of the
standardized Laplacian matrix
${\widetilde L}_{k}(n)=\dfrac{1}{n}(kI-C-C^2-\ldots-C^k)$, where $C=[c_{uv}]$ is the matrix of a
cyclic permutation of order~$n$ with entries $c_{uv}=\left\{
\begin{array}{ll}
 1,  &v-u\in \{1,\;n-1\},\\
 0,  &\text{otherwise.}
\end{array} \right.$
The digraphs whose standardized Laplacian matrices are ${\widetilde L}_{k}(n)$
belong to the class of balances digraphs, which are relevant to the synthesis
of multiagent control protocols (see, e.\,g., [23]).
In balanced digraphs, the total weight of arcs converging to any vertex
is equal to the total weight of arcs diverging from it.
Besides that, for these digraphs, the total weight of converging
arcs is the same for all vertices, i.\,e., they are regular.

If $n$ is even and $k<n/2$, then of special interest are the matrices
of the form ${\widetilde L}_{k}(n)+S$, where $S=[s_{ij}]$ is the standardized Laplacian matrix
of a digraph obtained from the digraph with standardized Laplacian matrix $\dfrac{1}{n}(I-C^{n/2})$
by decreasing the weights of some arcs. It turns out that the maximum
imaginary part of the eigenvalues of ${\widetilde L}_{k}(n)+S$ does not exceed such a part for
${\widetilde L}_{k}(n)$, and the spectrum of ${\widetilde L}_{k}(n)+S$ is invariant under any change
of nonzero $s_{ij}$ and $s_{ji}$ that preserves $s_{ij}+s_{ji}$. In other words, the spectra of the
standardized Laplacian matrices ${\widetilde L}_{k}(n)+S$ depend on the total weights of opposite
``diagonal'' arcs of the digraph rather than on their individual weights.

\begin{theorem}
The boundary of the polygon $S(n)$ with vertices $(19)$
converges$,$ as $n\to\infty,$ to the curve made up by the
parts of two cycloids whose parametric equations are$:$
$\:z(\tau)=x(\tau)+\mathrm{i}\,y(\tau)\:$ and
$\:z(\tau)=x(\tau)-\mathrm{i}\,y(\tau),$ where
$\tau\in[0,2\pi]$ and
\begin{gather*}
x(\tau)=(2\pi)^{-1}(\tau-\sin\tau),\\
y(\tau)=(2\pi)^{-1}(1   -\cos\tau).
\end{gather*}
\end{theorem}

\begin{figure}[t]
\centering{\includegraphics[clip]{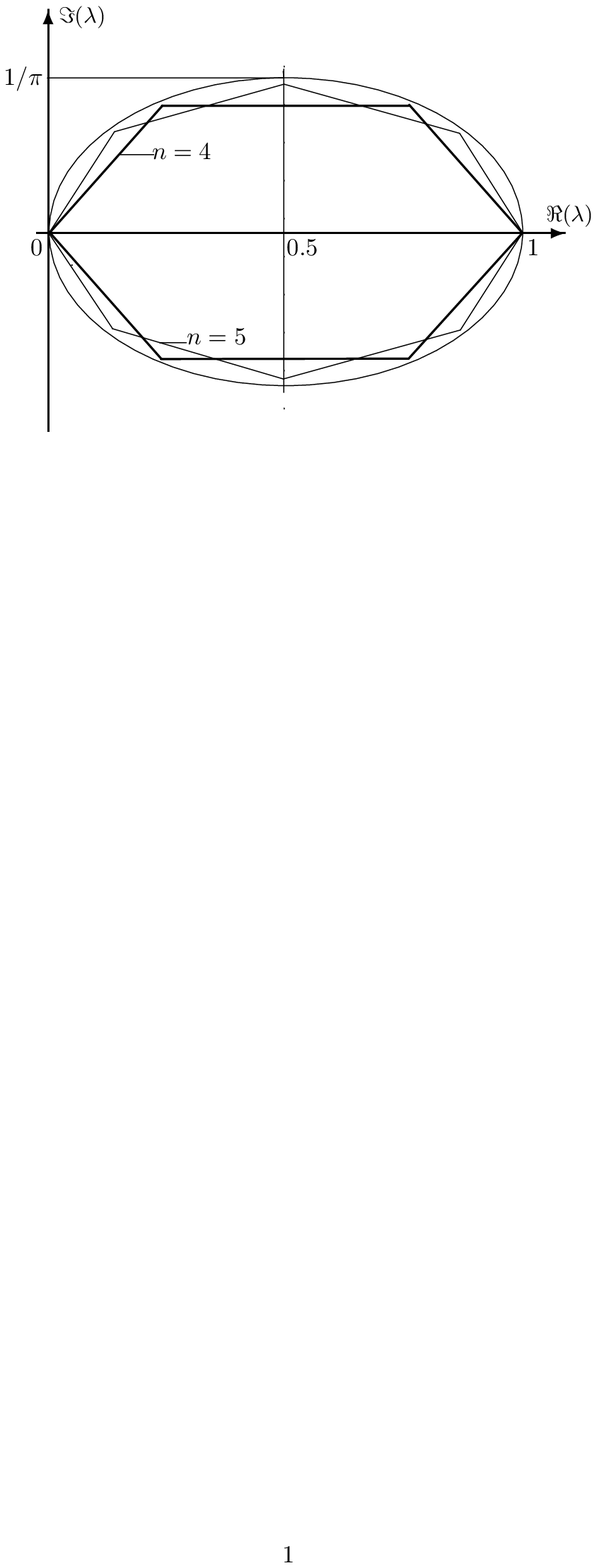}}
\caption{The polygon of the spectra of standardized digraph Laplacians
at $n=4$ and $n=5$ and the limit oval specified by Theorem~11.}
\bigskip
\end{figure}

Fig.~$2$ shows the polygons $S(n)$ at $n=4$ and $n = 5$
as well as the limit curve whose equation is given by Theorem~11.

As mentioned above, the results on localizing the spectra
of Laplacian matrices given in this section are necessary
for the analysis and synthesis of control algorithms that contain
reaching consensus as one of their elements (see, e.\,g., [31]).

\section{Conclusion}

Recently, decentralized control of multiagent systems has become one of the most popular
and rapidly evolving branches of control theory. Dozens of research groups have published
many hundreds of papers and since these groups are working concurrently and are studying
similar models, their results substantially overlap. To get familiar with this trend, the
recent surveys and monographs [23,\,32,\,37,\,38] can be recommended.
The methods of algebraic graph theory play an increasingly important part in the studies of
the last decade dealing with multiagent systems. More specifically, the subject of these
investigations is the relationship between the topological properties of digraphs that represent
information interchange among agents and the spectral properties of the corresponding Laplacian
matrices. At the same time, the Laplacian theory of directed graphs\footnote[6]{In [31] this area was
called an ``unexpected new mathematical territory.''} is still insufficiently developed; the need for
new strong results in this field is very acute. One more thesis accepted by most researchers is that
for a new stage of its development, this discipline requires more experimental
studies\footnote[7]{They are still scarce; see, e.\,g., [39, 40].} and more practical
applications of the theoretical results. These will reveal to what extent the proposed
algorithms are robust to perturbations and discrepancies between theory and real behavior.

\revred{B.T. Polyak}

\received{21.07.2008}
\end{document}